\documentclass[12pt]{article}
\usepackage{caption}
\usepackage[utf8]{inputenc}
\usepackage{amsmath,amsthm,bm,amsfonts,mathrsfs,mathtools,hyperref,lipsum}
\usepackage[numbered]{bookmark}
\usepackage{mdframed}
\usepackage{courier}
\usepackage{lineno}
\usepackage{authblk}
\hypersetup{
    colorlinks,
    citecolor=black,
    filecolor=black,
    linkcolor=black,
    urlcolor=black
}
\usepackage[a4paper, total={7in, 10in}]{geometry}
\usepackage{graphicx, float, multicol, balance}
\usepackage[version=4]{mhchem}
\usepackage[sorting=none, maxnames=5, style=nature]{biblatex}
\usepackage{listings}
\title{Temperature-Enhanced Coercive Field by Chiral Molecules}

\author[1]{Yael Kapon$^\#$}
\author[1]{Lilach Brann$^\#$}
\author[1]{Shira Yochelis}
\author[2]{Jonas Fransson}
\author[3]{Dimitar D. Sasselov}
\author[1]{Yossi Paltiel}
\author[4]{S. Furkan Ozturk\thanks{ozturk@caltech.edu}}

\affil[1]{Department of Applied Physics, The Hebrew University, Jerusalem 9190401, Israel}
\affil[2]{Department of Physics and Astronomy, Uppsala University, Uppsala 752 36, Sweden}
\affil[3]{Department of Astronomy, Harvard University, Cambridge, MA 02138, USA}
\affil[4]{Division of Geological and Planetary Sciences, California Institute of Technology, Pasadena, CA 91125, USA}

\date{\today}

\begin{document}
\maketitle

\makeatletter\def\Hy@Warning#1{}\makeatother
\def\thefootnote{\#}\footnotetext{These authors contributed equally to this work.}\def\thefootnote{\arabic{footnote}}

\begin{abstract}

The chiral-induced spin selectivity (CISS) effect demonstrates a strong coupling between electron spin and molecular chirality, enabling spin-controlled interactions between chiral molecules and magnetic surfaces. While CISS experiments have revealed robust changes in the spin-polarization properties of magnetic materials upon chiral molecular adsorption, the temperature dependence of these effects remains poorly understood. Here, we investigate the temperature dependence of the chirality-induced increase in magnetic coercivity by ribose-aminooxazoline (RAO) crystals on ferromagnetic surfaces. RAO was selected as a conglomerate-forming, thermodynamically stable crystalline chiral organic molecule with prebiotic relevance that has previously been shown to induce strong spin-dependent changes in magnetic minerals. Contrary to classical expectations that magnetic coercivity weakens at elevated temperatures, we observe a significant increase in magnetic coercivity ($\approx$1mT over a 60$^\circ$C temperature change) with increasing temperature. These results support a vibronic contribution to CISS arising from electron–phonon interactions and demonstrate that spin-dependent interactions between chiral molecules and magnetic surfaces can become more effective at higher temperatures, providing new insight into the microscopic origins of CISS and the environmental robustness of spin-controlled asymmetric processes. 
\end{abstract}

\begin{figure}[H]
    \centering
    \includegraphics[width=0.7\textwidth]{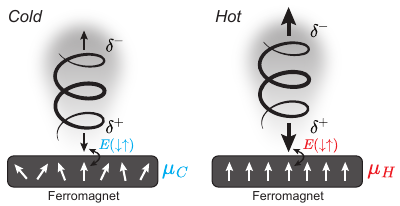}
    \caption[Intro]{Chiral molecules can alter the magnetic properties of ferromagnetic surfaces through spin-exchange interactions, arising from transient spin splitting in a chiral potential due to the chiral-induced spin selectivity (CISS) effect. Spin polarization across a chiral molecule due to CISS is higher at higher temperatures, leading to an increase in chirality-induced magnetization and enhanced ferromagnetic order at higher temperatures: $\mu_H > \mu_C$.}
    \label{Intro}
\end{figure}

The chiral-induced spin selectivity (CISS) effect\textemdash is an empirical phenomenon revealing a strong interaction between electron spin and molecular chirality, even at room temperature \cite{naaman2012chiral, naaman2019chiral, bloom2024chiral}. The strong spin-chirality coupling due to CISS \cite{ziv2019afm, safari2024enantioselective} allows for achiral magnetic minerals to function as chiral reagents for chemical reactions, which can be controlled by electron spin \cite{banerjee2018separation, bloom2020asymmetric, metzger2021dynamic}. Concurrently, the reciprocal nature of this effect was demonstrated by showing that chiral molecules can induce magnetization on magnetic thin films \cite{ozturk2023magnetization, kapon2023magneto,ben2017magnetization} and locally change the magnetic anisotropy of thin films \cite{sun2024colossal,moharana2025chiral}(\autoref{Intro}a). Although various independent measurements confirm CISS-related effects, a complete microscopic understanding of CISS remains an ongoing challenge \cite{evers2022theory}. Notably, recent theoretical work by Fransson emphasizes the role of electron correlations in explaining the high levels of spin-polarization observed in experiments \cite{fransson2019chirality}. 

Additionally, the effect of temperature on CISS through electron-phonon interactions presents an intriguing complexity \cite{bloom2024chiral, kim2023chiral, fransson2020vibrational}, with experimental results showing trends that appear incompatible with each other \cite{das2022temperature, qian2022chiral, alwan2023temperature, yang2023real}. This complexity is partly due to the fact that temperature also affects the material and molecular properties involved, complicating efforts to isolate the CISS component of the observed effects. To address these challenges, we examined the temperature dependence of chirality-induced avalanche magnetization in chiral ribose-aminooxazoline (RAO) crystals on magnetic surfaces stable within our temperature range (\autoref{Intro}). RAO is a conglomerate-forming central RNA and DNA precursor of prebiotic relevance, characterized by high thermodynamic stability and a well-established crystal structure, and it exhibits strong spin-dependent interactions with magnetic surfaces \cite{springsteen2004selective, patel2015common, powner2009synthesis, xu2020selective, ozturk2023crystallization}.

In contrast to classical expectations for spin-controlled processes, our experimental results and theoretical analysis show that the adsorption of chiral molecules leads to a significant increase in magnetic coercivity—via their spin-selective interactions with the magnetic surface—that becomes more enhanced at higher temperatures (\autoref{MOKE}). This supports a vibronic contribution to CISS, indicating that phonon-assisted couplings play a significant role in enhancing the ferromagnetic order due to the CISS effect \cite{fransson2025chiral}. These findings shed light on the temperature dependence of CISS, which remains an active area of research with seemingly incompatible findings. 

\section*{Results}

We studied the effect of temperature on chirality-induced magnetization using a magneto-optical Kerr effect (MOKE) microscope and showed an increase in magnetic coercivity with increasing temperature.

MOKE microscopy is an optical technique used to measure the surface magnetization of a material by reflecting polarized light from its surface, producing image contrast that is directly proportional to the relative magnetization \cite{kapon2023magneto, ozturk2023magnetization, sharma2020control}. MOKE microscopy was used to measure the chirality-induced avalanche magnetization on a ferromagnetic surface by enantiopure RAO crystals.

\begin{figure}[H]
    \centering
    \includegraphics[width=\textwidth]{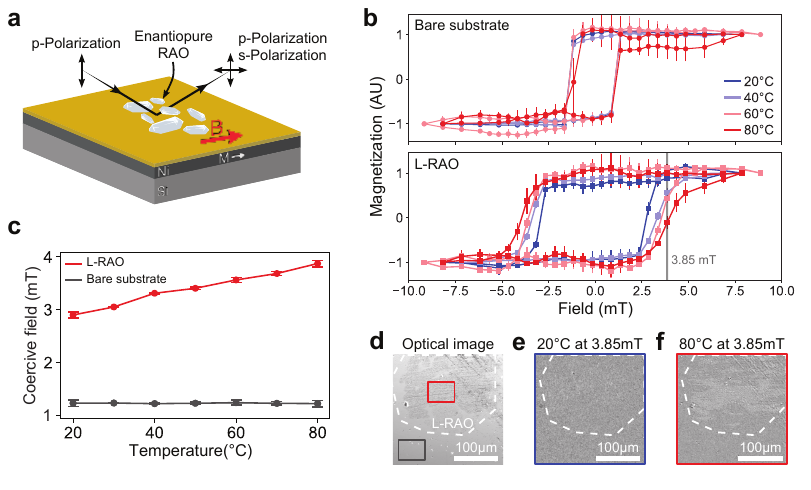}
    \caption[MOKE]{\textbf{a} Enantiopure RAO crystals were drop-cast onto a Si/Ni(50 nm)/Au(5 nm) thin film, forming a sample imaged with a MOKE microscope. In this setup, linearly polarized light is used to image the surface magnetization of the sample. The rotation of the polarization of the incident light corresponds to the surface magnetization. \textbf{b} Magnetic hysteresis loops are obtained by sweeping the magnetic field and averaging the intensity of regions far from (top) and beneath (bottom) homochiral \textit{L}-RAO crystals. As the temperature increases, the loops beneath the crystals widen, indicating an increased coercive field, while the bare surface’s coercive field remains constant. The vertical line at 3.85 mT is chosen to illustrate the difference in magnetization images at different temperatures, as seen in panels \textbf{e} and \textbf{f}. \textbf{c} Coercive fields as a function of temperature for regions beneath (red) and far from (gray) the \textit{L}-RAO crystals. The coercive field beneath the crystals is 2 mT higher at room temperature and increases linearly with rising temperature. \textbf{d} Optical image of the sample. A dashed white line outlines the area covered by \textit{L}-RAO crystals, and regions beneath the crystals and far from the crystals, where magnetization loops were taken, are marked in red/gray, respectively. \textbf{e} Magnetic image at 20°C and a 3.85 mT magnetic field, showing all domains flipped. \textbf{f} Magnetic image at 80°C and a 3.85 mT magnetic field, where some domains beneath the crystals remain unflipped.}
    \label{MOKE}
\end{figure}

Enantiopure RAO crystals were formed on a Si/Ni(50 nm)/Au(5 nm) thin film by drop-casting an RAO water solution, as illustrated in \autoref{MOKE}a. A commercial Evico-Magnetics GmbH Magneto-optical Kerr effect (MOKE) microscope, equipped with an electromagnet oriented parallel to the sample surface, was used to image the local magnetic effects induced by the enantiopure RAO crystals. The MOKE microscope provides qualitative imaging of surface magnetization by detecting the Kerr rotation, which reflects the change in the linear polarization of light as a function of the local magnetization (\autoref{MOKE}a). The contrast in the MOKE images corresponds to regions with different magnetization directions relative to the incident light, allowing us to assess the influence of the enantiopure RAO crystals on adjacent magnetic domains. It is important to note that MOKE microscopy here is not used for measuring induced magnetization upon adsorption of RAO molecules, but the coercive field of the sample with and without the RAO molecules, i.e., the sample's response to an outer magnetic field. 

\begin{figure}[H]
    \centering
    \includegraphics[width=\textwidth]{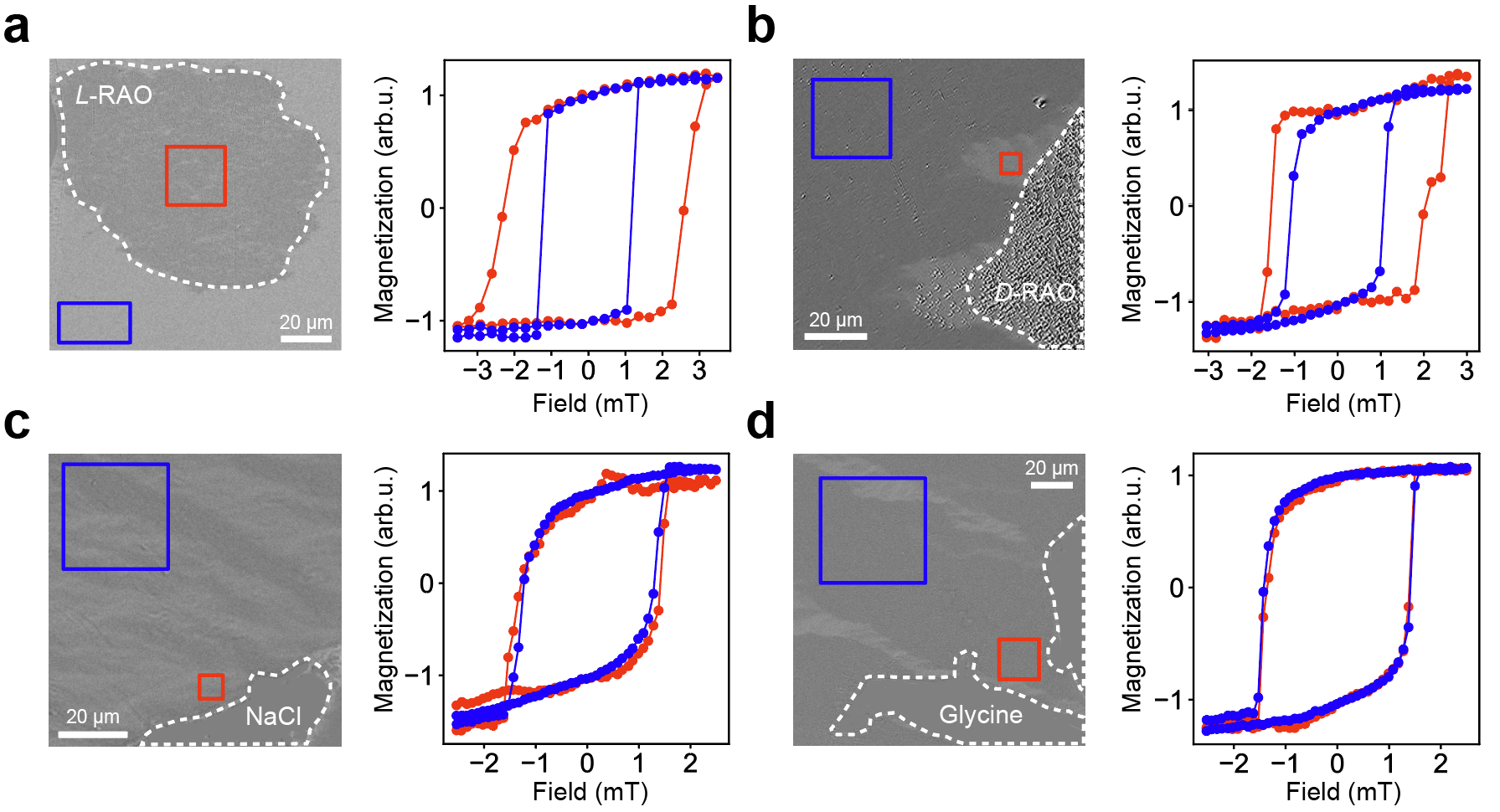}
    \caption[AchiralvsChiral]{Kerr hysteresis loops of \textbf{a} \textit{L}-RAO, \textbf{b} \textit{D}-RAO, \textbf{c} glycine, and \textbf{d} NaCl on the Ni/Au substrate. Achiral crystals on the magnetic substrate do not lead to any change in the magnetic properties, whereas both enantiomers of RAO, a chiral crystal, lead to a significant increase in the coercive field. Figure adapted from Ozturk, S.F., Bhowmick, D.K., Kapon, Y. et al. Chirality-induced avalanche magnetization of magnetite by an RNA precursor. Nat. Commun. 14, 6351 (2023). Licensed under a Creative Commons Attribution 4.0 International License (http://creativecommons.org/licenses/by/4.0/).” }
    \label{AchiralvsChiral}
\end{figure}

Surface magnetization images of the Ni film around the enantiopure RAO crystals at different magnetic fields and temperatures (20°C blue, 40°C light blue, 60°C light red, 80°C red) were taken. Magnetic hysteresis loops were obtained from surface magnetization images by sweeping the magnetic field and averaging the intensity in areas far (\autoref{MOKE}b top) and under (\autoref{MOKE}b bottom) from the enantiopure RAO crystals within the same image. As the temperature increases, the magnetic hysteresis loops under the crystals widen, indicating an increase in the coercive field. In contrast, the coercive field of the bare magnetic surface far from the crystals remains constant across all temperatures. During the measurements, RAO crystals remain stable at all temperatures. Racemic RAO was not studied since RAO crystallizes as a conglomerate, inevitably creating homochiral domains. Achiral crystals of glycine and sodium chloride were studied under the same conditions as control experiments (\autoref{AchiralvsChiral}). Unlike the chiral samples, they showed no change in coercivity relative to the bare substrate, further confirming that the observed effect arises from molecular chirality.

Starting with a negative field that saturates the sample, a uniform dark magnetization is observed, corresponding to -1 magnetization. As the magnetic field increases, the Ni magnetization flips to bright (+1 magnetization) at 1 mT; however, it requires an additional 2 mT to flip the domains surrounding the RAO crystals. When sweeping the field back down, the domains far from the crystals flip at -1 mT, while those under the crystals flip at -3 mT. A video of the MOKE images during a single sweep is presented in Videos S1 and S2, and an example of the resulting images is presented in Figure S3. The coercive field is plotted as a function of temperature in \autoref{MOKE}c for regions under (red) and far (gray) from the crystals. The coercive field under the crystal is 2 mT higher than that of the bare surface far from the crystals and increases linearly with temperature. 

The optical image of the sample is shown in \autoref{MOKE}d with the area of the crystals marked in a dashed white line. The areas under the crystals (red) or far from the crystals (gray) from which the magnetization loops were taken are marked on the image. Corresponding magnetic images at a 3.85 mT magnetic field are shown in \autoref{MOKE}e (20°C) and \autoref{MOKE}f (80°C). In this field, all the domains flipped at 20°C, while at 80°C, some light domains are left under the crystals. Both enantiomers of RAO were studied, and they led to similar phenomena of increased magnetic coercivity, as presented in Supporting Information Section 1.2.

At higher temperatures, a more gradual transition from one magnetization to the other under the crystals is seen by the increasingly slow transition toward magnetic saturation in the magnetic hysteresis loops in \autoref{MOKE}b. This behavior likely arises from reduced domain size at higher temperatures. Consequently, although the coercive field is stronger, the extent of avalanche magnetization decreases (Supporting Information Section 1.5), highlighting a tradeoff between the strength of induced magnetization and its spatial spread with increasing temperatures.

\section*{Discussion}

\subsection*{Chiral molecules resemble Carnot engines}

How can we account for the increased effectiveness of CISS with temperature without involving microscopic details? We will start with an intuitive thermodynamic picture and incorporate microscopic analyses later in the discussion.

\begin{figure}
    \centering
    \includegraphics[width=\textwidth]{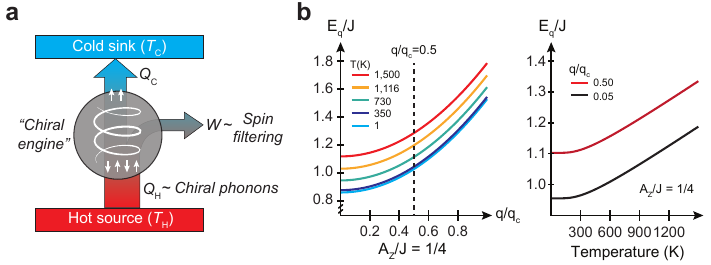}
    \caption[Discussion]{\textbf{a} The temperature dependence of CISS can be interpreted with a thermodynamic picture. In this model, the work done by chiral molecules—reducing entropy by filtering accessible spin states—is compensated by heat supplied to the molecules in the form of phonons, thereby enhancing the efficiency of this “chiral engine.” \textbf{b} Magnon energy dispersion relation as a function of (left) coupling strength between spin and vibrational modes, q, for five different temperatures between 1 K and 1,500 K and (b) temperature for two different values of q. Here, the spin-lattice coupling $A_z=1/4$, axial anisotropy $I_0=-1/8$ in units of $J=80$ meV.}
    \label{Discussion}
\end{figure}

CISS relates to the spin-filtering of electrons, meaning that chiral molecules reduce entropy by decreasing the number of accessible microstates of the electronic spin distribution at the molecule–surface interface.

We first consider the electronic spin distribution as the system of interest. To leading order, electron number and energy are conserved during spin filtering, allowing this subsystem to be described within a microcanonical ensemble. According to the first law of thermodynamics,

\[
dE = TdS + dW = 0 .
\]

Spin filtering reduces the number of accessible spin microstates, implying $dS < 0$ for the electronic subsystem. This entropy reduction requires positive work to be performed by the chiral molecules.

We next consider the chiral molecules as a separate subsystem interacting with their surroundings. Since the entropy of the electronic subsystem decreases, the entropy of the chiral molecules must increase according to the second law of thermodynamics. The free energy change of the molecular subsystem is

\[
dG = dH - T dS .
\]

Because $dS > 0$ for the molecules, the $-T dS$ term becomes increasingly negative at higher temperatures. Since the spin filtering process is not spontaneous, heat must be supplied to the molecular subsystem to maintain a positive free energy cost.

In this picture, chiral molecules operate as Carnot-like engines, absorbing heat from their surroundings and converting it into work that reduces the entropy of the electronic spin distribution (\autoref{Discussion}a). Consequently, these chiral engines operate more effectively when heat is provided.

This thermodynamic picture is an intuitive way of visualizing the temperature dependence of CISS. However, the effect itself is a complex phenomenon involving spin-orbit coupling, electron correlations, and electron spin-phonon interactions. Fransson investigated the role of phonons and vibronic interactions in CISS, constructing a microscopic theory of CISS beyond the Born–Oppenheimer approximation. This model elucidates the exchange of angular momentum as electrons traverse a helical trajectory \cite{fransson2020vibrational, fransson2022charge}, which is introduced below and elaborated upon in our complementary theoretical study \cite{fransson2025chiral}. Our experimental results align with this microscopic framework, indicating that phonon-chirality couplings, amplified by temperature, play an essential role in CISS.

\subsection*{Microscopic model}

Having provided an intuitive thermodynamic picture, we will now present a microscopic model connecting phonons to electron spin, and thereby to magnetization. This model produces the curves displayed in \autoref{Discussion}b.

The strength of the coercive field is a measure of the axial magnetic anisotropy, which corresponds to the energy barrier stabilizing a ferromagnet against spin excitations. These spin excitations, or magnons, are essentially destructive to the ferromagnetic order. However, by studying the magnon spectrum, we gain insight into the cause of the temperature-enhanced coercive field observed in the experiments (\autoref{MOKE}).

The anisotropic Heisenberg model describes the magnetic excitations for a collection of interacting spin moments $S_i$ distributed on a two-dimensional lattice at the coordinates $r_i$. In Hamiltonian form, this model can be written as:

\begin{equation}
    H_H = -J\sum_{ij}\textbf{S}_i\cdot\textbf{S}_j-I\sum_{ij}S^z_i S^z_j
\end{equation}

where the first term accounts for isotropic interactions between the spins, while the second provides an axial anisotropy, resulting in a preferred orientation for the total magnetic moment. The interaction parameters $J$ and $I$ are chosen to be positive, such that a ferromagnetic ground state is favored.

This model is capable of describing the ferromagnetic state of a local moment ferromagnet for temperatures below the Curie temperature, $T_C$. While the Heisenberg interaction $J$ defines a ferromagnetic ground state, this state is spatially isotropic, with the effect that a magnetic moment cannot be derived solely in terms of this interaction. This is also a consequence of the Mermin and Wagner theorem, which states that local isotropic interactions cannot give rise to long-range order for systems with spatial dimensions less than three \cite{mermin1966absence}. The anisotropic contribution $I$ relaxes this fundamental principle and introduces an energy gap such that any excited state of the spin system requires additional energy to be occupied—energy that may be provided by the temperature of the surrounding environment. Eventually, this energy gap is closed by the thermal energy at sufficiently high temperatures, leading to the destruction of the ferromagnetic order. This is the reason for the destruction of ferromagnetism with temperature.

The unexpected increase in magnetic coercivity with temperature observed in this work can be explained by the coupling between nuclear vibrations and the localized spin moment \cite{fransson2017microscopic}. This is the origin of the enhanced chirality-induced magnetization with increasing temperatures.

This interaction is captured by the addition of

\begin{equation}
    H_{SQ} = -\sum_{ij}\textbf{S}_i \cdot \mathbb{A}_{ij} \cdot \textbf{Q}_j
\end{equation}

where $Q_i$ denotes the spatial displacement of the nucleus at $r_i$, while $\mathbb{A}_{ij}$ represents a coupling between the localized spin moment at $S_i$ and the displacement $Q_j$ (or the nucleus at $r_j$). The nuclear displacements result in a field acting on the localized spin moments, resembling that of a magnetic field; hence, we may refer to it as a pseudo-magnetic field. It should be noted that $\mathbb{A}_{ij}$ is a tensorial interaction with the form of a susceptibility, which arises from charge-spin fluctuations in the compound. While this susceptibility can be considered negligible in inversion-symmetric crystal structures, $\mathbb{A}_{ij}$ is non-vanishing in chiral structures \cite{fransson2023chiral}, where inversion symmetry is broken by definition. Introducing this component to the model is relevant in the current context, where we have chiral molecules adsorbed on a magnetite surface.

As we are interested in the effect of the spin-lattice coupling on the magnetic anisotropy, we consider the linear spin wave spectrum that results from the introduced model. The details are presented in a complementary work \cite{fransson2025chiral}, and here, we merely present the salient features from this investigation.

In the absence of spin-lattice coupling, at lower temperatures, the linear spin wave spectrum in the limit of long wavelength is captured by the energy dispersion relation $E_q = \alpha q^2 + \beta$, where $q$ is the reciprocal of the wavelength, $\alpha > 0$ is related to the exchange energy $J$, and $\beta > 0$ represents the anisotropic energy in terms of $I$. Therefore, the physics of the composed system is dominated by processes that are very well captured within this theoretical picture.

At higher temperatures, the dominant processes that govern the physics are related to the spin-lattice coupling, such that the energy dispersion relation must be replaced by $E_q(T)=\alpha q^2 + \beta + \gamma T$, where the additional anisotropic contribution $\gamma$ provides a linear temperature dependence of the energy dispersion, as seen in the plotted curve in \autoref{Discussion}b. The derivation of this linearity is discussed in \cite{fransson2025chiral} and in the SI in section 4. The effect of this linear contribution to the anisotropy energy leads to an increase in the coercive field with temperature, in agreement with the experimental observations (\autoref{MOKE}). Physically, we interpret this addition to the anisotropic energy as a new channel that diverts the energy away from exciting spin waves in the system and instead generates lattice vibrations. This term is connected to heat supplied to the chiral engine, as we explained in the thermodynamic picture.

The combination of these two terms—one being destructive and the other constructive to the thermodynamic order—stabilizes the system of chiral molecules on a magnetic surface against thermal fluctuations. This provides persistent magnetization necessary for observing spin-controlled asymmetric processes due to CISS in natural environments. This inherent stability of our mechanism, together with RAO's superior thermodynamic stability \cite{powner2010phosphate}, enhances our scenario’s environmental plausibility.

\subsection*{Increased coercivity and the role of magnetic substrate}

The magnetic response in our system manifests primarily as an increase in coercive field, rather than as a reversal of the net magnetization. This behavior differs from earlier reports of magnetization switching in perpendicular-anisotropy stacks \cite{ben2017magnetization}, which are distinct from both our in-plane Ni/Au films and bulk magnetite.

The MOKE loops remain symmetric in our measurements and show no evidence of exchange bias, though we cannot fully exclude subtle effects below our detection threshold. This observation supports a scenario of anisotropy change rather than molecule-driven magnetization switching. A possible explanation is that the molecules induce local magnetic moments at the surface \cite{alwan2021spinterface}. Such surface effects can alter magnetic anisotropy but may not be sufficient to reverse the magnetization of the ferromagnet within an exchange length.

To distinguish between coercivity enhancement and true magnetization switching, we performed complementary Hall effect measurements on Ni films with Au or oxide capping layers (Supporting Information Section 2). These measurements show an increase in Hall resistance upon RAO adsorption; however, the response lacks consistent directionality, likely due to nonspecific molecular binding at the surface. Therefore, based on our current measurements, we cannot conclude that RAO induces net magnetization switching in Ni/Au systems. However, our theoretical model allows us to probe the net magnetization induced by chiral molecules and its temperature dependence. It predicts that $M_z$, the net out-of-plane magnetization, decreases with increasing temperature \cite{fransson2025chiral}, unlike the magnetic coercivity, which is enhanced with temperature in agreement with our measurements.

Surface chemistry appears central to these considerations. On magnetite, the magnetic surface that was studied in connection to prebiotic chemistry of RAO, RAO directly binds through hydrogen bonding, a strong chemical interaction that brings the molecules into close proximity with the substrate. In contrast, on 5 nm gold-capped nickel, RAO interacts more weakly and less directly through electrostatic forces. This weaker physisorption also keeps the molecules farther from the surface, leading to reduced interfacial coupling and less organized dipole orientation. Molecular orientation is particularly important: as shown by Nguyen et al. \cite{nguyen2024mechanism}, oppositely oriented molecules can invert the direction of the CISS effect, and strong collective co-alignment of molecules is required for a strong CISS signal. 

Nickel surfaces introduce an additional complication. Their metallic nature induces a mirror potential that screens surface charges, effectively reducing the influence of adsorbed molecules on spin-dependent charge redistribution due to CISS \cite{liu2025dynamical}. This screening effect should be weaker on semiconducting magnetite, allowing stronger proximity interactions with chiral molecules.

In summary, RAO adsorption on magnetite produces a more pronounced CISS effect than on Ni/Au. This enhancement arises from (i) stronger chemical binding, (ii) stronger alignment of dipoles, (iii) closer molecular proximity to the surface, and (iv) weaker electronic screening by the semiconducting substrate.

Our earlier studies of RAO on magnetite, which revealed a handedness-dependent inversion of the circular dichroism signal, suggest that the RAO–magnetite interface can produce a net magnetization change within the magnetite exchange length (around 9 nm). In contrast, MOKE measurements of RAO on Ni/Au show no evidence of exchange bias in the hysteresis loops \cite{ozturk2023magnetization}. Taken together, these results clarify the subtle differences in the magnetic behavior of RAO on magnetite versus Ni/Au and provide a mechanistic basis for the enhanced CISS effect observed on magnetite.

Last but not least, it is also important to consider the role of magnetic domain size and structure. Our MOKE measurements demonstrate that the observed effects depend on magnetic domain size. In multidomain thin films, such as those studied here, in-plane magnetization switching requires a strong induced effective magnetic field, while coercivity can increase more readily through local changes in anisotropy. Under these conditions, any induced magnetization is transient and disappears once the molecules are removed. In contrast, systems with frustrated geometries—such as perpendicular-anisotropy stacks or single-domain particles—are more susceptible to stable magnetization reversal. Single-domain, low-coercivity magnetite particles represent the most suitable magnetic substrates for our system. It is therefore important to investigate whether RAO can induce permanent magnetization switching in these low-coercivity grains. These particles are also geochemically plausible under early Earth and Martian conditions and can be synthesized prebiotically from iron–carbonate solutions \cite{hurowitzetal2017, toscaetal2018}.

\section*{Conclusion}

Our findings show that spin-controlled asymmetric processes due to the CISS effect can locally increase the coercivity of a ferromagnetic thin film. Contrary to classical expectations, temperature, an important environmental variable, actually enhances this process. Our experiments reveal that chirality-induced coercivity by RAO becomes stronger as the temperature increases. These results not only provide insights into the temperature dependence of CISS, an active area of research, but also reinforce the stability of spin-controlled mechanisms involving CISS across various environmental conditions. The inherent stability of CISS, which distinguishes it from classical magnetic phenomena, arises from spin-phonon couplings, which appear to be crucial for the high degree of spin polarization observed in CISS experiments. Thus, our results offer both a deeper understanding of CISS and highlight its significance in the context of the origin of life's homochirality.

\section*{Methods}

\subsection*{Preparation of magnetic surfaces}

For the MOKE measurements, ferromagnetic Ti(8 nm)/Ni(30 nm)/Au(5 nm) surfaces were evaporated by electron beam under a high-vacuum of $3 \times 10^{-7}$  Torr on silicon (100) wafer. The thin gold coating was used to prevent the oxidation of the magnetic substrate and preserve its spin-polarization properties. The surfaces were cleaned thoroughly with acetone and ethanol before every experiment.

For the Hall measurements, Ni(40 nm)/Au(10 nm) surfaces were evaporated onto a Si/SiO wafer using electron beam evaporation (VST Metals Evaporator) under pressure of $3\times10^{-6}$ Torr. An aluminum mask was used to form four, $3\times3$mm\textsuperscript{2}, corner contact pads on the sample for better electrical connectivity to the surface. Ti(10 nm)/Au(50 nm) pads were evaporated under cooling conditions on top of the surface.

\subsection*{RAO synthesis}

\textit{D}- and \textit{L}-ribo-aminooxazolines (RAO) were synthesized by the reaction of two equivalents of cyanamide (5 g, 0.12 mol) with one equivalent of the \textit{D}- and \textit{D}-ribose (9 g, 0.06 mol) in aqueous ammonia (3.5\%, 10 mL). After the reaction, enantiopure RAO was crystallized, the solution was filtered, and the crystals were dried and ground into powder. 20 mM solutions of \textit{D}- or \textit{L}-RAO were prepared in pure water and used in the experiments.

\subsection*{RAO crystallization}

For the MOKE experiments, small (around 1-10 $\mu$m) and dense crystals were obtained by drop casting 5 $\mu$L of \textit{L}/\textit{D}-RAO solution on Ti(8 nm)/Ni(30 nm)/Au(5 nm) surfaces. Samples were then placed in the fridge (-18°C) for two cycles of two hours, with a two-hour interval at room temperature. Finally, samples were dried overnight at room temperature and areas of amorphous aggregation and dense crystallization were obtained. For the hall experiments, 10$\mu$L \textit{D}-RAO solution was drop-cast onto the Ni(40 nm)/Au(10 nm) surface and dried overnight at room temperature.

\subsection*{MOKE microscopy}

MOKE imaging and magnetometry were performed using a commercial Evico Magnetics GmbH magnetooptical Kerr microscope. The measurements were taken in the longitudinal configuration. An in-plane magnetic field was generated by an electromagnet obtained from the microscope supplier powered by a Kepco BOP 100-4DL power supply. For the optical imaging of substrates, 20X commercial Zeiss objective lenses were used. A piezo stage actively stabilized mechanical vibrations of the sample. Temperature was controlled by a Peltier element and was measured continuously using a Pt thermistor.

\section*{Acknowledgments}

We thank Naama Goren from the Hebrew University Applied Physics Department for her help in interpreting the Hall measurements and Maurice Saidian and Alexander Oginets from the Hebrew University Nano Center for Ni/Au evaporation. Y.P. acknowledges the funding from Marie Skłodowska-Curie Actions under Horizon Europe framework program (CISSE project No. 101071886), Carl Zeiss Stiftung (HYMMS project No. P2022-03-044), and U.S Air Force (grant No. FA8655-24-1-7390). Y.K. thanks the Israel Council of Higher Education's VTT fellowship for women in STEM. S.F.O. and D.D.S. acknowledge the Harvard Origins of Life Initiative for funding and its members for fruitful discussions. S.F.O. further acknowledges support from the Kavli-Laukien fellowship program, Caltech startup funds, the William H. Hurt Scholarship Program, and the Fellows of King's College Cambridge for their hospitality.

\printbibliography

\newpage
\clearpage
\setcounter{page}{1}
\setcounter{table}{0}
\setcounter{section}{0}
\setcounter{figure}{0}
\setcounter{equation}{0}
\bookmarksetup{startatroot}

\begin{center}
    \Large\textbf{Supporting Information}
    
\end{center}
\tableofcontents

\renewcommand{\thefigure}{S\arabic{figure}}
\setcounter{figure}{0}
\newpage

\section{MOKE Measurements}
\subsection{MOKE Setup}
MOKE measurements were taken by a commercial Evico Magnetics GmbH magneto-optical Kerr microscope equipped with an electromagnet and a piezo controller for mechanical stabilization, as seen in Fig. S1.
\begin{figure}[H]
\centering
\includegraphics{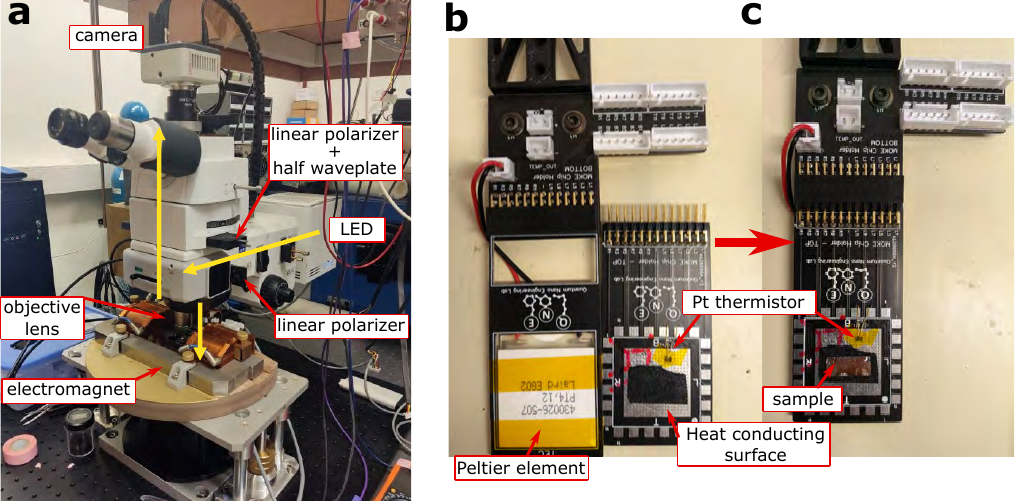}
\caption[MOKE setup]{\textbf{a} Magneto-optical Kerr effect microscope (MOKE) setup is displayed. The magnetic surface is imaged by a Zeiss objective, and an in-plane magnetic field was generated by an electromagnet placed around the imaging plane of the microscope. The yellow arrows represent the optical path. \textbf{b} A printed circuit board (PCB) with a Peltier element for heating connected to \textbf{c} a PCB sample holder. The sample is glued to a heat-conducting surface using carbon tape with a Pt thermistor to monitor temperature.}
\label{MOKE setup}
\end{figure}

\subsection{Effect of RAO Handedness}
\begin{figure}[H] 
\centering 
\includegraphics{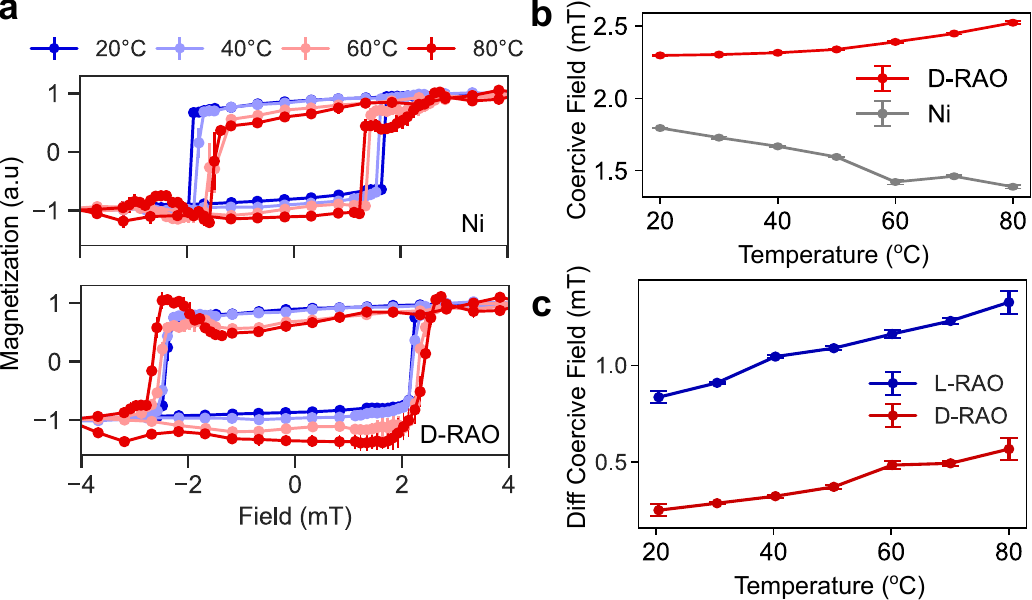} \caption[MOKE D]{\textbf{a} Magnetic hysteresis loops of Ni with D-RAO crystals, measured far from (top) and beneath (bottom) the crystals using MOKE. \textbf{b} Coercive field of Ni far from (gray) and beneath the crystals (red). \textbf{c} Coercive field differences due to L-RAO (blue) and D-RAO (red) crystals.} 
\label{MOKE D} 
\end{figure}

The MOKE experiment was repeated using both L- and D-enantiomers of RAO crystals. In \autoref{MOKE D}a, the magnetic hysteresis loops of Ni are presented for regions far from (top) and directly beneath (bottom) the D-RAO crystals. For the L-RAO sample, the magnetic properties of the bare Ni surface remained unchanged at all temperatures. In contrast, for the D-RAO sample, the coercive field of the bare Ni surface decreased at higher temperatures. Nonetheless, the difference in the coercive field between the regions beneath and far from the crystals increased with temperature, as shown in \autoref{MOKE D}b. In \autoref{MOKE D}c, this difference is compared for both L- and D-enantiomers, where a similar linear trend is observed for both.  The chiral molecules act as strong pinning sites with a mechanism that does not favor a particular direction, as previously discussed in previous works \cite{ozturk2023magnetization,kapon2023magneto}. In addition, it was previously observed that a-chiral crystals such as glycine and salt do not cause a pinning effect on the surface (see \autoref{AchiralvsChiral}), and therefore, we attribute the pinning effects seen here to chirality-induced magnetization and conclude that it does not have a favored magnetization direction in this experimental setup.

\subsection{Magnetic Field Sweep}
\begin{figure}[H] 
\centering 
\includegraphics{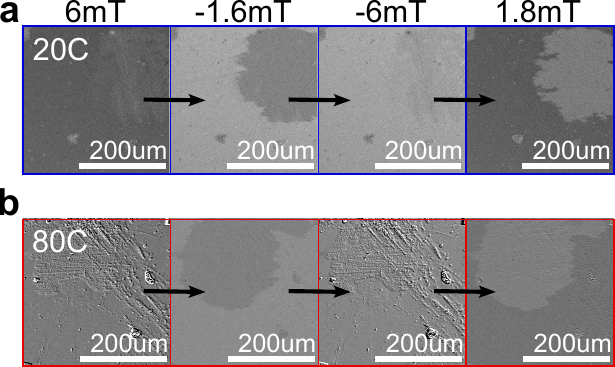} \caption[MOKE sweep]{MOKE images at different magnetic fields at 20C (\textbf{a}) and 80C (\textbf{b})} 
\label{MOKE sweep} 
\end{figure}

Magnetic hysteresis loops of Ni were measured by sweeping the magnetic field from positive to negative and back to positive. MOKE images at various magnetic fields are shown in \autoref{MOKE sweep}. Upon reaching the coercive field of bare Ni, the domains far from the crystals flip from dark to bright. A higher demagnetization field is required to flip the domains beneath the crystals. This process is symmetric and occurs similarly when sweeping the field from negative to positive. Full hysteresis measurements at 20°C and 80°C are available in Videos S1 and S2, respectively.  

\textbf{Video S1}: Kerr hysteresis measurement of the Ni/Au surface with RAO crystals at 20C. Domains around the chiral crystals have a higher magnetic coercivity, and they do not flip until a higher demagnetizing field is applied compared to the domains far from the chiral crystals.

\textbf{Video S2}: Kerr hysteresis measurement of the Ni/Au surface with RAO crystals at 80C. Domains around the chiral crystals have a higher magnetic coercivity, and they do not flip until a higher demagnetizing field is applied compared to the domains far from the chiral crystals.

\subsection{Sample Stability at Higher Temperatures}
\begin{figure}[H] \centering \includegraphics{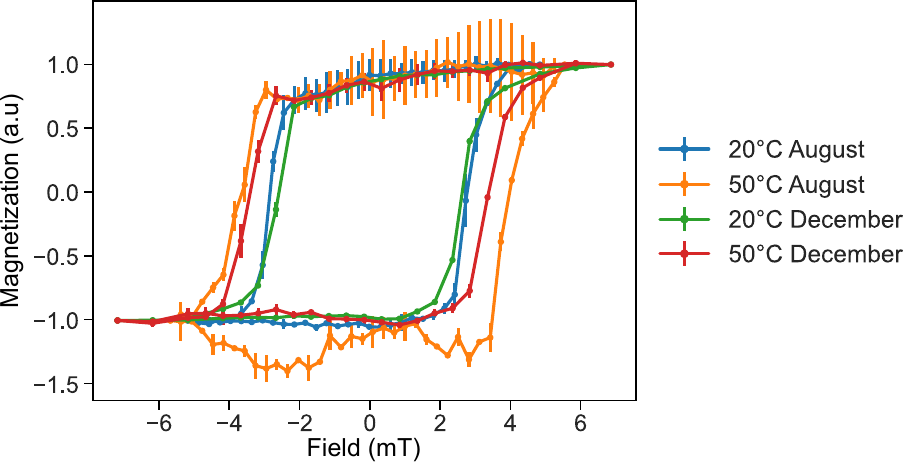} \caption[MOKE hysteresis loops]{MOKE hysteresis loops measured in August and December after multiple heating and cooling cycles. The magnetic properties remained constant.} \label{summer and winter} \end{figure}

MOKE hysteresis measurements were repeated after several heating and cooling cycles and over a period of months. As shown in \autoref{summer and winter}, the magnetic properties of the sample remained unchanged. This suggests that the crystals are stable at elevated temperatures and over time. The observed local increase in the coercive field at higher temperatures in Figure 2 is therefore attributed to the coupling between nuclear vibrations and the localized spin moment, rather than any conformational changes in the molecules.

\subsection{Domain Area Analysis}

\begin{figure}[H] \centering \includegraphics{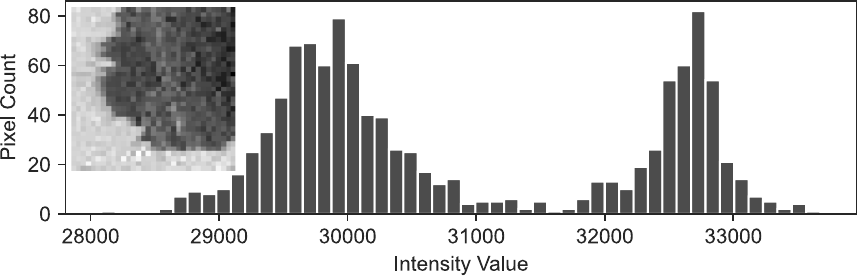} \caption[Intensity Histogram]{Intensity histogram from a MOKE image (inset), showing two Gaussian distributions for the light and dark domains.} \label{intensity histogram} \end{figure}

The relationship between domain size, magnetic field strength, and temperature was investigated. For each MOKE image, the flipped domain area was quantified by analyzing the intensity histogram. Once the domains began to flip, the intensity histogram displayed two Gaussian distributions corresponding to the light and dark domains. An example histogram and the associated magnetic image are shown in \autoref{intensity histogram}. The image was binned to help differentiate between light and dark pixels and reduce noise.

\begin{figure}[H] \centering \includegraphics{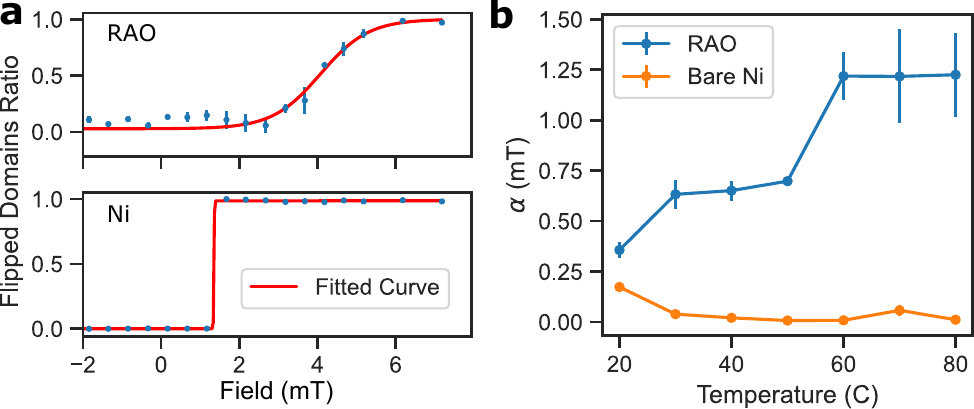} \caption[Domain Area Analysis]{Relative flipped domain area as a function of magnetic field beneath the crystals (\textbf{a} top) and far from the crystals (\textbf{a} bottom). A $\tanh(x)$ fit is shown in red. \textbf{b} Fitting parameter $\alpha$, representing the slope of the transition, as a function of temperature beneath (blue) and far from (orange) the crystals.} \label{domain area analysis} \end{figure}

A threshold value (e.g., 31,500 in this example) was defined between the two distributions, with pixels above the threshold classified as bright and those below classified as dark. The ratio of bright pixels to the total pixel count in the image was calculated and plotted as a function of the magnetic field for regions beneath the crystals (\autoref{domain area analysis}a top) and far from the crystals (\autoref{domain area analysis}a bottom).

The data were fitted using the function $A \tanh\left(\frac{x - B}{\alpha}\right) + C$, where the parameter $\alpha$ represents the slope of the transition. In the absence of the crystals, a smaller absolute value of $\alpha$ was observed, indicating a sharper transition and faster domain flipping. \autoref{domain area analysis}b presents $\alpha$ as a function of temperature, comparing areas beneath and far from the crystals. At higher temperatures, smaller domain sizes resulted in slower transitions and reduced magnetization spread.

\subsection{Errors in MOKE Measurements}

MOKE measurements were repeated at least five times to minimize errors arising from vibrations and sample drift. Since MOKE imaging involves subtracting a magnetically saturated background image, it is particularly sensitive to movement and drift, which were more pronounced at higher temperatures. This increased instability is evident in Video S2 compared to S1. Error bars in the hysteresis loop represent a standard error of the mean ($\frac{\sigma}{\sqrt{n}}$; where $\sigma$ is the standard deviation).

\subsection{MOKE Analysis Code}
The code used for data processing and analysis is provided below:

\begin{lstlisting}
# MOKE Analysis

# Import the data:
def readfile(file, path): 
    """
    Read and preprocess data from a file.
    """
    df = pd.read_csv(path + '\\' + file, sep='\t')
    data = np.array(df)
    # Normalize, center, and flip the hysteresis loop
    data[:, 1] = data[:, 1] - (np.max(data[:, 1]) + np.min(data[:, 1])) / 2  # Center
    data[:, 1] = data[:, 1] / np.max(data[:, 1])  # Normalize
    data[:, 1] = -data[:, 1]  # Flip
    return data

# Extract hysteresis loop from images
def hist_on_area(path, file, X, Y):
    """
    Extract hysteresis loop signal from images.
    """
    data = readfile(file, path)
    field = data[:, 0]
    signal = np.zeros(len(field))
    cnt = 0
    
    for images in sorted(os.listdir(path)):
        if images.startswith("image0"):
            im = Image.open(path + "\\" + images)
            im_array = np.asarray(im.rotate(0))  # Rotate and convert to array
            signal[cnt] = np.sum(im_array[X[0]:X[1], Y[0]:Y[1]])
            cnt += 1

    # Correct drift and normalize
    drift = signal[-1] - signal[0]
    signal = signal - np.linspace(0, drift, num=len(signal))  # Drift correction
    signal = signal - (np.max(signal) + np.min(signal)) / 2  # Center
    signal = signal / signal[np.argmax(field)]  # Normalize
    return field, signal

# Calculate coercive field width
def find_width(x, vector):
    """
    Calculate the width of the hysteresis loop.
    """
    sign_change_indices = [i for i in range(len(vector) - 1) if np.sign(vector[i]) != np.sign(vector[i + 1])]
    x1, y1, m1, b1 = interpolate_zero(x, vector, sign_change_indices[0])
    x2, y2, m2, b2 = interpolate_zero(x, vector, sign_change_indices[1])
    return abs(-b2 / m2 - -b1 / m1)

# Helper for linear interpolation
def interpolate_zero(x, y, idx):
    y1, y2 = y[idx], y[idx + 1]
    x1, x2 = x[idx], x[idx + 1]
    m = (y2 - y1) / (x2 - x1)
    b = y1 - m * x1
    return x1, y1, m, b

# Domain Size Analysis
def bin_image(image_array, bin_size):
    """
    Reduce noise by binning the image.
    """
    height, width = image_array.shape
    cropped_array = image_array[:height // bin_size * bin_size, :width // bin_size * bin_size]
    shape = (height // bin_size, bin_size, width // bin_size, bin_size)
    return cropped_array.reshape(shape).mean(axis=(1, 3))

def find_ratio(full_path, X, Y, threshold, bin_size):
    """
    Calculate the ratio of bright pixels in a specified region.
    """
    im = Image.open(full_path).crop((X[0], Y[0], X[1], Y[1]))
    binned_array = bin_image(np.asarray(im), bin_size)
    bright = np.sum(binned_array > threshold)
    return bright / binned_array.size

# Sigmoidal Fitting
def sigmoidal(x, Xc, A, alpha, Yc):
    """
    Sigmoidal model based on a hyperbolic tangent function.
    """
    return A * np.tanh((x - Xc) / alpha) + Yc

def find_fit(x, y, p0=[0.1, 1, 0.01, 0.5]):
    """
    Fit data to the sigmoidal function.
    """
    params, covariance = curve_fit(sigmoidal, x, y, p0=p0)
    return params, covariance
\end{lstlisting}

\section{Hall Measurements}

When an electric current flows through a conductor in a perpendicular magnetic field, the Lorentz force creates a transverse Hall voltage ($V_H$). In ferromagnets, the anomalous Hall effect introduces an additional term from intrinsic magnetization. The total Hall voltage is: 
$V_{H} = R_{0} B + R_{s} M$.
Where $R_{0} B$ is the regular Hall term and $R_{s} M$ is the anomalous term due to the material's magnetization. The anomalous Hall effect is commonly used to probe surface magnetization induced by chiral molecules \cite{kumar2017chirality,eckshtain2016cold}.

\subsection{Hall Setup and Methods}
Hall measurements were performed on three types of samples: Ni(40 nm)/Au(10 nm), Ni(7mn)/Au(7nm) patterned hall bar, and Ni(7nm)/Au(7nm)/Al2O3(1nm) patterned hall bar.
Three types of molecules were studied: D-RAO, L-RAO, and a-chiral Glycine.
The measurements were conducted under no magnetic field or a constant external magnetic field generated by an out-of-plane electromagnet powered by a Kepco BOP 100-4DL power supply.  The current was supplied by Keysight B2901A precision source while perpendicular Hall voltage was measured in parallel by Keitley 2400 SourceMeter. Temperature was controlled by a Peltier element and was measured continuously using a Pt thermistor. The current was scanned from -1 mA to 1 mA and back and forth. These measurements were repeated 5 times for each direction and at different temperatures: 30°C, 40°C, 50°C, 60°C, 70°C, and 80°C.

\begin{figure}[H] \centering \includegraphics{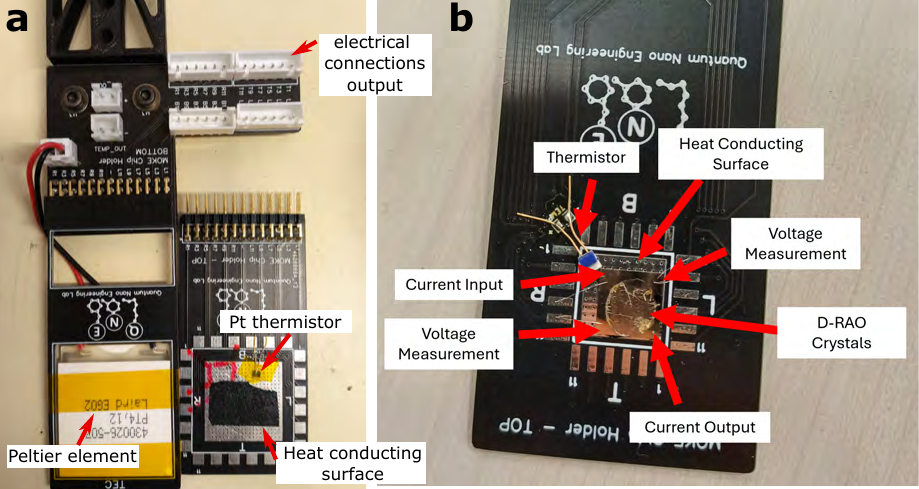} \caption[Hall setup]{\textbf{a} A printed circuit board (PCB) with a Peltier element for heating connected to \textbf{b} a PCB sample holder. The Ni/Au sample is glued to a heat-conducting surface using carbon tape with a Pt thermistor to monitor temperature. Au pads are evaporated to the corners of the sample. The pads are electrically connected to the PCB sample holder by gold wires, which in turn are electrically connected within the PCB to the electrical connection output.} \label{Hall setup} \end{figure}

\subsection{Analysis of Hall Measurements}

\begin{figure}[H] \centering \includegraphics{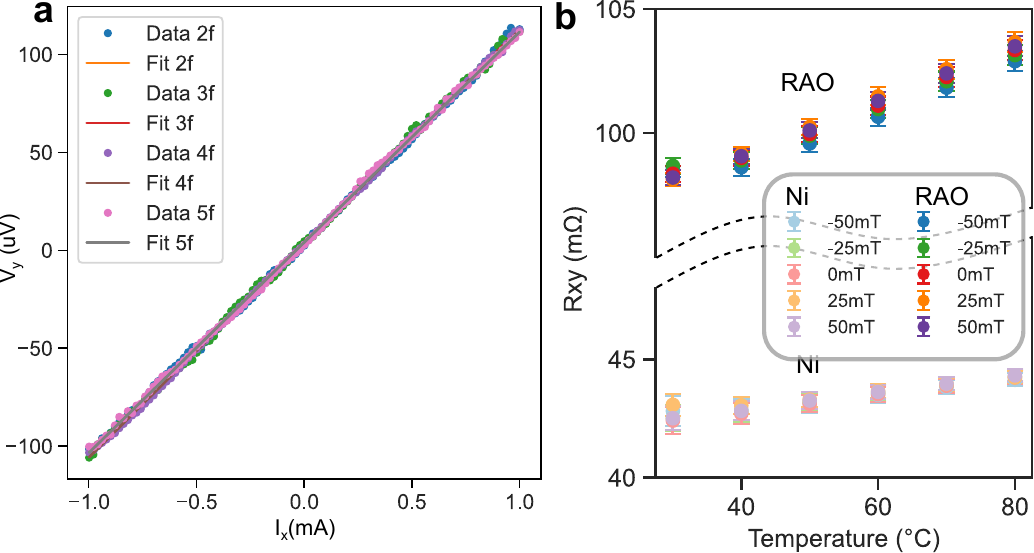}\caption[Hall data]{\textbf{a} Hall IV curve for Ni/Au with RAO crystals at 50mT and 80°C \textbf{b} $R_{xy}$ for all the temperatures (30-80°C) and fields (-50 mT - blue,-25 mT - green,0 mT - red,25 mT - orange,50 mT - purple) with (dark) and without (light) RAO.} \label{Hall data} \end{figure}

A linear relation was observed for each measurement of the Hall voltage as a function of the current. A linear function was fitted to each measurement, with the slope of the fit being the hall resistance,$R_{xy}$. For example, the five repetitions in the forward direction of the Hall IV curve and the corresponding fits for Ni/Au with RAO crystals at 50mT and 80°C are shown in \autoref{Hall data}a. $R_{xy}$ was averaged across all repetitions for each specific external field, temperature, and direction (forward or reverse), both with and without the D-RAO molecules. For each of these averages, a standard error was calculated. $R_{xy}$ for all the temperatures and fields are represented in \autoref{Hall data}. 
As shown, the behavior is similar for all magnetic fields, including those with negative values. We expected to observe a negative Hall voltage for the negative fields, but this was not the case. This, along with the relatively small external fields applied (up to 50 mT), suggests that the effect we measured is the Anomalous Hall Effect, and in the applied field range, the magnetization of the Ni did not change.

To observe the changes to $R_{xy}$ after molecular deposition, we calculated the change in percentages as: $\Delta R_{xy} = (R_{xy}^{molecules}-R_{xy}^{substrate})/R_{xy}^{substrate} \times100$
The error is the standard error of mean between measurements.

\subsection{Inconclusive Sign for Hall Effect Measurements}

\begin{figure}[H]
    \centering
    \includegraphics[width=\textwidth]{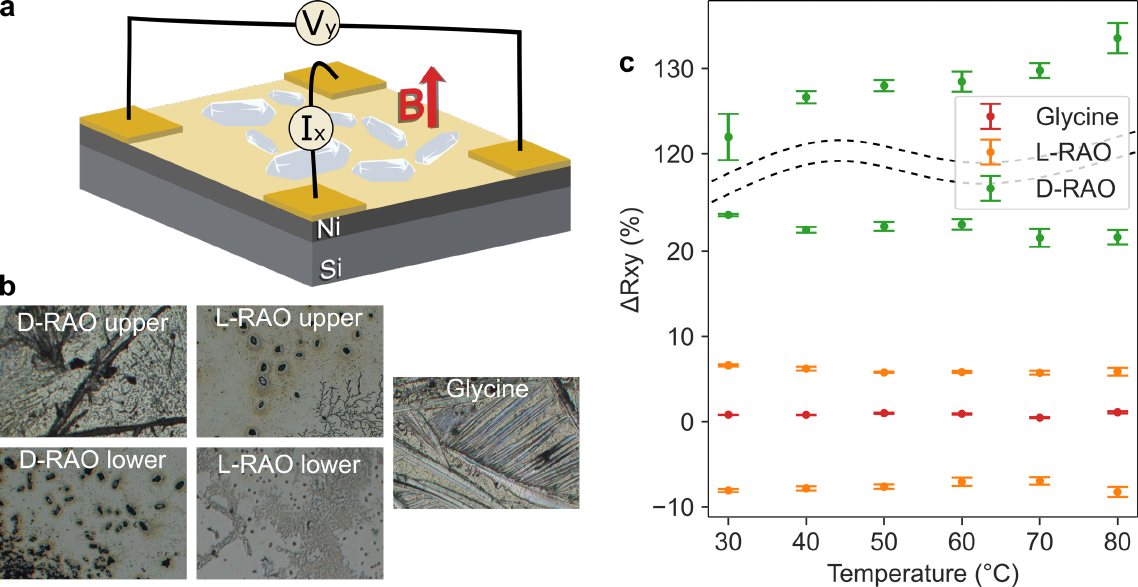}
    \caption[Hall]{\textbf{a} Using a simple Hall configuration on an $8 \,\text{mm} \times 8 \,\text{mm}$ sample, the Hall voltage ($V_{y}$) was measured perpendicular to the applied current ($I_{x}$) on a Ni/Au substrate, with and without RAO crystals, under an external magnetic field of $-25$ mT. \textbf{b} Optical images of the crystals formed on different samples, showing significant variations in crystal morphology. \textbf{c} Temperature-dependent change in Hall resistance ($\Delta R_{xy} (\%)$) for D-RAO (green), L-RAO (orange), and glycine (red). $R_{xy}$ is significantly larger in the presence of RAO compared to glycine, but the directionality differs between samples.}
    \label{Hall}
\end{figure}

The magnetization induced by RAO crystals on a Si/Ni(40 nm)/Au(10 nm) thin film was measured using a simple Hall bar configuration, as shown in \autoref{Hall}a. Au pads were fabricated at the corners of the Si/Ni(40 nm)/Au(10 nm) film. A current ($I_{x}$) was applied, and the perpendicular Hall voltage ($V_{y}$) was measured under an external magnetic field of $-25$ mT applied normal to the surface. Enantiopure D-RAO, L-RAO, and achiral glycine were drop-cast onto the substrate and left to dry. Hall voltages were measured before and after molecule adsorption to avoid variations due to device geometry or intrinsic magnetic properties between samples. Two samples were prepared with each RAO enantiomer and one with glycine. As shown in the optical images in \autoref{Hall}b, the crystals formed on the surface varied significantly between samples. 

The Hall resistance ($\Delta R_{xy} (\%)$) was extracted following the protocol described in the previous section and is presented in \autoref{Hall}c. In all chiral samples, $R_{xy}$ changes significantly after molecule deposition, whereas the achiral sample shows no change. However, both the magnitude and sign of the change vary between samples of the same chirality. Moreover, the measurements do not exhibit a clear temperature dependence. Because of this variability, we cannot determine whether magnetization reversal occurs due to the molecules. This is likely because Au does not provide specific binding to RAO, leading to non-uniform molecular orientation both within and between samples. It has previously been shown that molecular orientation on a surface can invert preferential spin conductance \cite{nguyen2024mechanism}. The crystalline structure provides directional facets relative to the surface during nucleation, but these facets may differ between samples and across large crystal domains, as in this case.

\subsection{Sequential Deposition}

\begin{figure}[H]
    \centering
    \includegraphics[width=\textwidth]{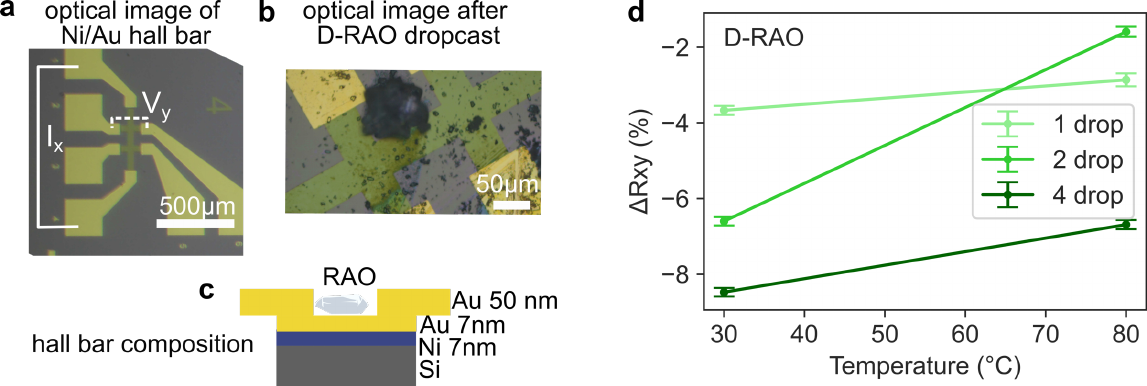}
    \caption[Hall]{\textbf{a} Optical image of the Ni/Au Hall bar device. \textbf{b} Optical image of the RAO crystals after four sequential depositions. \textbf{c} Schematic of the Hall device structure. \textbf{d} Change in Hall resistance ($\Delta R_{xy}$) at $30^\circ$C and $80^\circ$C as a function of the number of deposited drops.}
    \label{drops}
\end{figure}

To address the variability observed in large-area samples, we performed a sequential deposition experiment using a fabricated Hall bar device. This geometry reduces sample-to-sample variation and ensures that crystal growth proceeds from existing nucleation sites, thereby preserving the same crystalline facets. The Hall bar was fabricated using standard photolithography with a Ni(7 nm)/Au(7 nm) channel and Au (50 nm) contact pads, as illustrated in \autoref{drops}c. A current was applied through the channel and the transverse Hall voltage was measured, with the device shown in the optical image in \autoref{drops}a. 

Sequential deposition was carried out by drop-casting $1 \,\mu\text{L}$ of 20 mM D-RAO solution onto the Hall bar. After each deposition, the Hall signal was recorded. In total, one, two, and four drops were deposited, with the final crystal morphology shown in \autoref{drops}b. The resulting changes in Hall resistance ($\Delta R_{xy}$) at $30^\circ$C and $80^\circ$C are shown in \autoref{drops}d. The Hall signal increases systematically with each additional drop, although the overall magnitude of the effect remains small. As in the larger-area devices, no clear temperature dependence is observed; however, this method allows us to resolve a consistent directionality within a single sample under controlled chemical conditions.

\subsection{Oxide Capping}

\begin{figure}[H]
    \centering
    \includegraphics[width=\textwidth]{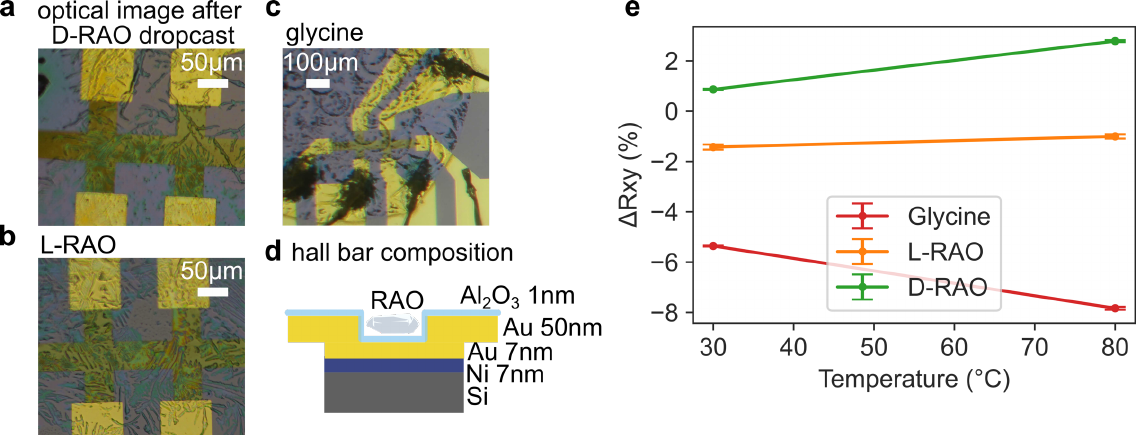}
    \caption[Hall]{\textbf{a–c} Optical images of D-RAO, L-RAO, and glycine crystals deposited on Al$_2$O$_3$-capped Hall bars. \textbf{d} Schematic of the Hall device structure. \textbf{e} Change in Hall resistance ($\Delta R_{xy}$) after molecular deposition.}
    \label{oxide}
\end{figure}

To create a more directional surface chemistry, we evaporated a 1 nm Al$_2$O$_3$ capping layer onto a Hall bar fabricated as described in the previous section (\autoref{oxide}d). The Hall signal was measured before and after deposition of 40 mM D-RAO, L-RAO, and glycine solutions (optical images in \autoref{oxide}a–c). The RAO samples produced a more uniform crystal coverage under these conditions. To achieve this, the RAO solutions were supersaturated by dissolving 40 mM of D- or L-RAO in water at $60^\circ$C until fully dissolved, followed by drop-casting onto the Hall bar while still hot. Crystals then formed upon cooling. Both D- and L-RAO showed more uniform crystallization compared to the uncapped case, whereas glycine did not.

The corresponding $\Delta R_{xy}$ values are shown in \autoref{oxide}e. Both RAO enantiomers enhanced the Hall signal slightly and in opposite directions, while glycine also induced a noticeable change. As in previous measurements, no clear temperature dependence was observed. However, in this configuration, the chemical interactions are more controlled. The weak overall signal is attributed to the increased separation between the ferromagnetic layer and the molecules, since the additional oxide layer reduces interfacial coupling.

Overall, we are unable to conclude that the RAO molecules induced any net magnetization. The main limitation lies in the nonspecific surface chemistry of the RAO–Au interface, in contrast to more established systems such as thiol-functionalized polypeptides that form strong Au–S bonds, or RAO–magnetite interactions stabilized through hydrogen bonding. In our case, the RAO molecules are likely positioned farther from the substrate and lack directional binding, which reduces the possibility of generating a robust net magnetization. Thus, even if local magnetization changes occurred, they would not necessarily translate into a measurable exchange bias at the Ni/Au surface.

\subsection{Hall Analysis Code}
The code used for data processing and analysis is provided below:

\begin{lstlisting}
# Hall Data Analysis
def import_data_from_file(file_path):
    """
    Import Hall data starting from the marker $$$.
    """
    marker = '$$$'
    start_row = 0
    with open(file_path, 'r') as file:
        for line_number, line in enumerate(file):
            if marker in line:
                start_row = line_number + 1
                break
    return pd.read_csv(file_path, sep='\t', skiprows=start_row)

# Organize Hall data
def divide_by_outer_loop(df, column_name):
    """
    Divide data based on unique values in the outer loop.
    """
    return {f"df_{value}": df[df[column_name] == value] for value in df[column_name].unique()}

def split_to_repetitions(df, chunk_size):
    """
    Split data into repetitions of a specified size.
    """
    return {f"{i//2+1}{'f' if i%2==0 else 'b'}": df.iloc[i:i + chunk_size] for i in range(0, len(df), chunk_size)}

def find_fit_data(xdata, ydata, init_guess, low_bound, high_bound):
    """
    Fit Hall data to a linear model.
    """
    return curve_fit(linear_fun, xdata, ydata, p0=init_guess, bounds=(low_bound, high_bound))

def linear_fun(x, a, c):
    return a * x + c
\end{lstlisting}

\section{Linearity of the Energy Gap}

The interaction given in Eq.~(2) was derived in Ref.\cite{fransson2017microscopic} and is based on electronically mediated interactions between localized spin moments and nuclear displacement. This formulation is in direct analogy with both indirect spin-spin interactions as well as the electronic contribution to the atomic force constant. The interaction tensor $A_{ij}$ reflects the susceptibility associated with charge-spin fluctuations and is nonzero whenever the mediating electrons are either spin-polarized and/or spin-orbit coupled. In the present setup, both requirements can be presumed to be satisfied.

The close to linear temperature dependence of the dispersion relation is obtained from the Hartree approximation of the correction to the dispersion relation (see, e.g., Eq.~(7) in Ref.\cite{fransson2025chiral}). The temperature-dependent factor in this correction can be written as
\begin{equation}
    \delta E_q (T) \sim \frac{1}{\beta} 
    \ln \left( 
    \frac{1 - e^{-2M\beta (I_0 - \alpha p_c^2)}}{1 - e^{-2M\beta I_0}}
    \right),
    \label{eq:deltaE}
\end{equation}
where $1/\beta = k_B T$ is the thermal energy (with $k_B$ the Boltzmann constant and $T$ the temperature), $M$ is the magnetic moment per atom, $I_0$ is the unperturbed spin anisotropy, and $\alpha p_c^2$ is the high-energy cut-off for the magnon band. 

For high temperatures ($\beta \to 0$), the exponentials can, to a good approximation, be Taylor expanded to first order in $\beta$. In this limit, the logarithm becomes independent of temperature, such that
\begin{equation}
    \delta E_q (T) \propto \frac{1}{\beta} \sim T.
\end{equation}
Hence, in the parameter range relevant to the present context, the correction is close to linear for $T > 250 \, \text{K}$.

\end{document}